\documentclass[12pt]{article}
\usepackage{diss}

\title{Feynman-Kac Derivatives Pricing on the Full Forward Curve}
\author{Kevin Mott\thanks{Rotman School of Management, University of Toronto. Email: \href{mailto:kevin.mott@rotman.utoronto.ca}{kevin.mott@rotman.utoronto.ca}. This paper is based on a chapter of my doctoral dissertation. I am especially grateful to Burton Hollifield for his mentorship throughout this project. I also thank my dissertation committee---Steve Spear, Chester Spatt, Tetiana Davydiuk, and Deeksha Gupta---for their support and feedback. For helpful comments and discussions, I thank Martin Larsson and Zahra Ebrahimi.}}
\date{\monthyear\today}

\begin{document}

\maketitle

\begin{abstract}
\noindent This paper introduces a no-arbitrage, Monte Carlo-free approach to pricing path-dependent interest rate derivatives. The Heath-Jarrow-Morton model gives arbitrage-free contingent claims prices but is infinite-dimensional, making traditional numerical methods computationally prohibitive. To make the problem computationally tractable, I cast the stochastic pricing problem as a deterministic partial differential equation (PDE). Finance-Informed Neural Networks (FINNs) solve this PDE directly by minimizing violations of the differential equation and boundary condition, with automatic differentiation efficiently computing the exact derivatives needed to evaluate PDE terms. FINNs achieve pricing accuracy within 0.04 to 0.07 cents per dollar of contract value compared to Monte Carlo benchmarks. Once trained, FINNs price caplets in a few microseconds regardless of dimension, delivering speedups ranging from 300,000 to 4.5 million times faster than Monte Carlo simulation as the state space discretization of the forward curve grows from 10 to 150 nodes. The major Greeks—theta and curve deltas—come for free, computed automatically during PDE evaluation at zero marginal cost, whereas Monte Carlo requires complete re-simulation for each sensitivity. The framework generalizes naturally beyond caplets to other path-dependent derivatives—caps, swaptions, callable bonds—requiring only boundary condition modifications while retaining the same core PDE structure.

\vspace{1em}
\end{abstract}

\newpage
\doublespacing

\section{Introduction}

The Heath-Jarrow-Morton (HJM) framework \parencite{heath1992bond} is the most general arbitrage-free approach to pricing interest rate derivatives, specifying stochastic evolution for the entire forward curve simultaneously and nesting earlier short-rate models as special cases. Under risk-neutral valuation, a derivative with contract features $\Xi$ (including expiry $T$) and payoff at expiry $h(T,f;\Xi)$ depending on the forward curve $f(t,T)$ has time-$t$ price given by the stochastic expectation:
\begin{equation}
    V(t) = \E\left[\exp\left(-\int_t^T r(s)\d s\right) h(T,f;\Xi)  \right]
\end{equation}
where $r(t)=f(t,t)$ is the short rate. Path dependence enters through two channels: the discount factor requires integrating the short rate $r(s)$ along the entire stochastic path of the forward curve from $t$ to $T$, and the payoff $h(T,f;\Xi)$ depends on the forward curve realized at expiry. This path dependence prevents closed-form analytical solutions for most derivatives, forcing practitioners to rely on numerical methods. Since the forward curve is infinite-dimensional, achieving tight estimates of derivative prices entails evaluating this expectation over a fine partition of the forward curve. Monte Carlo simulation of thousands or millions of stochastic forward curve paths becomes increasingly computationally expensive as this grid grows finer, forcing practitioners to choose between accuracy and computational feasibility. Suppose you simulate a contract and obtain a price with reasonable accuracy. If the counterparty changes the terms—even marginally adjusting strike, maturity, or other contract terms—you must resimulate entirely from scratch. Worse, computing Greeks requires perturbing contract parameters or underlying rates and drawing fresh paths for each sensitivity, making real-time risk management computationally expensive. 

This paper introduces \emph{Finance-Informed Neural Networks} (FINNs) to solve this computational bottleneck. FINNs circumvent Monte Carlo entirely by exploiting the Feynman-Kac theorem, which establishes that the stochastic expectation above satisfies a deterministic PDE that can be solved directly via deep learning. The approach combines two key insights: first, the Feynman-Kac transformation eliminates Monte Carlo simulation by replacing the stochastic pricing problem with a PDE characterization; second, neural networks trained via automatic differentiation solve high-dimensional PDEs efficiently, avoiding the curse of dimensionality that plagues traditional finite difference methods. Once trained, FINNs price derivatives in microseconds regardless of state space dimension, delivering speedups of many orders of magnitude over Monte Carlo. Critically, the major Greeks—theta and curve deltas—come for free, since they appear directly in the PDE being minimized during training. Other Greeks require only negligible additional computation via automatic differentiation. A further practical advantage is the approach's flexibility: regardless of the specific derivative contract—caplets, swaptions, callable bonds, or exotic path-dependent structures—the core PDE governing prices remains identical. Pricing different instruments requires only adjusting the contract features in the state variable and modifying the boundary condition to reflect the appropriate payoff function, without altering the fundamental PDE structure.

The methodology builds on \emph{physics-informed neural networks} (PINNs) pioneered by \textcite{raissi2019physics}, which embed governing differential equations directly into the loss function. Neural networks are universal function approximators whose evaluation cost scales gracefully with dimension. Modern deep learning frameworks compute exact derivatives via automatic differentiation, enabling direct evaluation of PDE residuals at any point in the state space. The neural network parameterizes the pricing functional, and training minimizes PDE violations alongside boundary condition penalties—critically, enforced without any forward simulation by evaluating terminal payoffs directly on the cross-section of historical forward curves rather than simulating curves to maturity.

I demonstrate this methodology by pricing interest rate caplets—call options on future LIBOR rates and fundamental building blocks for caps and swaptions. Using daily U.S. Treasury forward curve data from \textcite{gurkaynak2007us}, I estimate volatility via PCA, derive the caplet pricing PDE, and train neural networks across eight discretization levels ranging from $K=10$ to $K=150$ tenor points. The training procedure exploits the analytical zero-strike caplet solution as an additional anchor, disciplining the pricing functional with exact solutions where available. All results use local volatility (scaling with the square root of forward rate levels) to capture the empirical phenomenon that interest rate volatility scales with rate levels. Validated against a test set of 1,000 randomly sampled contracts, the trained FINNs achieve pricing accuracy within 0.04\textcent\ to 0.07\textcent\ compared to Monte Carlo benchmarks (per dollar of contract value), while requiring only consumer-grade hardware (8GB GPU) and evaluating in a few microseconds once trained. The computational advantage is dramatic: as I grow the state space (discretization of the forward curve) from 10 to 150 nodes, FINNs price caplets 300,000 to 4,500,000 times faster than Monte Carlo simulation. 

\section{Literature Review}

This paper sits at the intersection of term structure modeling, computational PDE methods, and machine learning for derivatives pricing.


Early short-rate models \parencite{vasicek1977equilibrium, cox1985theory, hull1990pricing} specify dynamics for the instantaneous rate and derive the entire term structure from this single state variable, offering analytical tractability at the cost of restricting all interest rate movements to a single factor. For interest rate derivatives, practitioners frequently use the formula of \textcite{black1976pricing} for caps and floors, which applies the Black-Scholes framework directly to forward rates. However, the Black model prices each caplet independently without specifying how forward rates co-evolve, potentially admitting arbitrage when pricing portfolios of caplets at different maturities.

\textcite{heath1992bond} revolutionized the field by modeling the entire forward curve simultaneously, yielding an arbitrage-free framework that nests earlier short-rate models as special cases. The cost of this generality is computational: the forward curve is infinite-dimensional, requiring discretization for numerical implementation. Subsequent work has taken two approaches to manage this complexity. One strand imposes structure to achieve dimensionality reduction, either through Markovian restrictions \parencite{cheyette2001markov} or through discretely-compounded forward rate specifications like the LIBOR market model \parencite{brace1997market}, which prices caps and swaptions directly in terms of observable market rates. My approach takes a different path, avoiding dimensionality reduction entirely—the neural network accommodates the full discretized forward curve without requiring Markovian structure or parametric restrictions. The Musiela parameterization \parencite{musiela2005martingale} reformulates HJM dynamics in terms of time-to-maturity rather than calendar time, a change of variables essential for PDE-based pricing methods.

PDEs have been understood to be at the heart of option pricing at least since the seminal work of \textcite{black1973pricing} and \textcite{merton1971theory}. Recent work applies neural networks to solve pricing PDEs in continuous-time models. \textcite{gopalakrishna2021aliens} and \textcite{zhang2022before} demonstrate neural network solutions for PDEs arising in macro-finance models, while a parallel literature focuses on learning optimal hedging strategies and/or derivatives prices. \textcite{buehler2019deep} pioneer the ``deep hedging'' framework, using deep reinforcement learning to hedge portfolios of derivatives under market frictions such as transaction costs and liquidity constraints, demonstrating that neural networks can approximate optimal hedging strategies in high-dimensional settings. Subsequent work \parencite{chen2023variational,yuana2024hedging,cao2020neural,cao2021deep,cao2023gamma} extends these ideas to volatility surface generation, barrier option hedging, and gamma/vega hedging with distributional reinforcement learning. \textcite{Fan23022026} use neural networks to parameterize drift and volatility functions in SDE models, training via stochastic gradient descent for European options and Kolmogorov PDEs for American options, and evaluate out-of-sample generalization on S\&P index and single-stock options. In the equity context, \textcite{huge2020differential} introduce differential machine learning, which augments neural network training with pathwise sensitivity estimates to improve pricing and hedging accuracy; \textcite{glasserman2025differential} identify biases in this approach for options with discontinuous payoffs and propose corrected sensitivity estimators for digital and barrier options.

A distinct methodological strand uses neural networks to \emph{solve} PDEs without discretizing the state space. \textcite{raissi2019physics} pioneered physics-informed neural networks (PINNs), which embed the governing PDE directly in the loss function via automatic differentiation, eliminating the need for spatial grids. This mesh-free approach scales naturally to high dimensions. \textcite{han2018solving} develop the deep BSDE method for backward stochastic differential equations, demonstrating solutions to parabolic PDEs in hundreds of dimensions, while \textcite{sirignano2018dgm} propose the Deep Galerkin Method as an alternative variational approach. For financial applications specifically, \textcite{beck2021solving} extend these methods to Kolmogorov PDEs, solving backward in time from terminal conditions. 

\section{Heath-Jarrow-Morton Model Refresher}

What follows is a non-rigorous refresher of the Heath-Jarrow-Morton \parencite{heath1992bond} paper. 

Consider a continuous trading interval $[0,\overline T]$ for a fixed $\overline T>0$ and probability space $(\Omega,\mathcal F,Q)$ where $\Omega$ is the state space, $\mathcal F$ is the $\sigma$-algebra representing measurable events, and $Q$ is a probability measure. Augmented, right-continuous complete filtration $\{F_t: t \in [0,\overline T]\}$ generated by $N\geq1$ independent Brownian motions $\{W_n(t):t\in[0,\overline T]\}_{n=1}^N$ initialized at zero. Let $\mathbb{E}[\cdot]$ denote expectations with respect to the probability measure $Q$. 

There exists a continuum of default-free pure discount bonds trading with differing maturities $T\in[0,\overline T]$. The price at time $t$ of a bond maturing at $T$ for all $T\in[0,\overline T]$ and $t\in[0,T]$ is denoted $P(t,T)$. Face values are normalized to 1: $P(T,T)=1$ for all $T\in[0,\overline T]$. Additionally, $P(t,T)>0$ and $\partial \log P(t,T)/\partial T$ exists for all $T\in[0,\overline T]$ and $t\in[0,T]$.\footnote{Normalize payoffs, no arbitrage, and forward rates are well-defined, respectively.}

Define the instantaneous forward rate at time $t$ for date $T>t$ as:
\begin{equation}
    f(t,T) = -\frac{\partial \log P(t,T)}{\partial T} \text{ for all } T\in[0,\overline T], \; t\in[0,T].
\end{equation}
Bond prices can be expressed in terms of forward rates: 
\begin{equation}
    P(t,T) = \exp\left(-\int_t^T f(t,s) \d s\right)  \text{ for all } T\in[0,\overline T], \; t\in[0,T]. \label{bondprice}
\end{equation}
The instantaneous forward rate is termed the \emph{spot rate} and is given by 
\begin{equation}
    r(t) = f(t,t) \text{ for all } t \in [0,\overline T].
\end{equation}

HJM starts with forward rate dynamics given as:
\begin{equation}
\d f(t,T) = \mu(t,T)\d t + \sigma(t,T) \cdot \d W(t). \label{hjm_0}
\end{equation}
Above, $\sigma(t,T)=\left(\sigma_n(t,T)\right)_{n=1}^N$ and $W(t) = \left(W_n(t)\right)_{n=1}^N$ are vectors in $\R^N$ representing the $n\geq1$ volatility processes. Both $\mu$ and $\sigma$ are assumed to be measurable, adapted, and integrable over $[0,T]$ almost everywhere with respect to $Q$. 
\begin{align*}
    \int_0^T \left|\mu(t,T)\right|\d t &< + \infty \\ 
    \int_0^T \sigma_n^2(t,T)\d t &< + \infty \text{ for } i = n,\ldots,N.
\end{align*}

Given a deterministic initial forward curve $f(0,T)$, the dynamics above uniquely determine the stochastic fluctuation of the entire forward curve according to 
\begin{equation}
    f(t,T) = f(0,T) + \int_0^t \mu(s,T)\d s + \sum_{n=1}^N \int_0^t \sigma_n(s,T)\d W_n(s)\label{evolution}
\end{equation}
for all  $0 \leq t \leq T$. 


Applying It\^o's Lemma to equation (\ref{bondprice}) and (\ref{hjm_0}) and applying the no-arbitrage restriction yields the familiar HJM result:\footnote{For a detailed derivation, see the original paper.}
\begin{equation}
    \mu(t,T) = \sigma(t,T)\cdot\int_t^T\sigma(t,s)\d s \label{HJM_1}
\end{equation}

This result implies that the dynamics of the entire forward curve are parameterized only by choice of $\sigma_i$ functions. The next section will discuss historical methods for doing so, as well as common methods for using this model for contingent claims valuation. The following section will discuss an improved contingent claims valuation approach, blending deep learning with the Feynman-Kac theorem. 

\subsection{Musiela Parameterization}
In computational applications, a change of variables termed the `Musiela Parameterization' is undertaken: from $(t,T)$ to $(t,\tau)$ where $\tau=T-t$ is termed the \emph{tenor} and represents time-to-maturity \parencite{musiela2005martingale}.
The model dynamics then become:
\begin{align}
     \d f(t,\tau) &= \mu(t,\tau) \d t + \sigma(t,\tau) \cdot \d W_t \\
     \d \tau &= - \d t
\end{align}
and the state space is $t \in [0,\overline T]$, $\tau \in [0,\overline T - t]$.

The no-arbitrage drift process is also amended:
\begin{equation}
    \mu(t,\tau) = \frac{\partial}{\partial \tau} f(t,\tau) + \sigma(t,\tau) \cdot \int_0^\tau \sigma(t,s)\d s \label{eq:musiela_drift}
\end{equation}

The Musiela form will be used in what follows for the rest of the paper.

\section{Standard Procedure for Computing the HJM Model}

The standard approach to implementing the HJM model involves three computationally straightforward steps: obtaining forward curve data, estimating the volatility structure via principal components analysis, and computing the no-arbitrage drift via numerical integration. All three steps are relatively inexpensive and can be performed once as preprocessing. I follow this standard procedure in my implementation, using widely available data and conventional estimation techniques.

The computational expense in HJM modeling does not arise from these preprocessing steps. Rather, it emerges when pricing path-dependent interest rate derivatives, which traditionally requires Monte Carlo simulation of thousands or millions of forward curve paths. This Monte Carlo bottleneck is the problem I address in subsequent sections by first casting the \emph{stochastic} simulation exercise into a \emph{deterministic} PDE and second by using FINNs to solve the pricing PDE directly.

\subsection{Data and Forward Curve Construction}
This paper uses daily instantaneous forward rate data constructed by \cite{gurkaynak2007us} from U.S. Treasury securities. The dataset provides fitted Svensson parameters $(\beta_0, \beta_1, \beta_2, \beta_3, \tau_1, \tau_2)$ estimated daily from all traded Treasury securities using the parametric yield curve methodology of \cite{svensson}. I utilize data from January 1, 2001 onward, chosen to focus on the modern interest rate environment while maintaining a sufficiently long time series for estimating volatility dynamics.

The Svensson model specifies the instantaneous forward rate as a function of tenor:
\begin{equation}
    f(\tau) = \beta_0 + \left(\beta_1 + \beta_2\frac{\tau}{\tau_1}\right) \exp\left(-\frac{\tau}{\tau_1}\right) + \beta_3  \frac{\tau}{\tau_2} \exp\left(-\frac{\tau}{\tau_2}\right)
\end{equation}

Raw Svensson parameter estimates occasionally exhibit extreme values due to market stress or thin trading in certain maturities. To ensure numerical stability, I filter observations using quantile-based outlier detection: for each parameter, I retain only observations between the 5th and 95th percentiles of its empirical distribution. This removes approximately 10\% of observations while preserving the full range of typical market conditions. For implementation, I evaluate the Svensson formula at a grid of equally spaced intervals over the interval $[0,5]$ years—appropriate for pricing short-maturity caplets. A final filtering step removes any dates where forward rates are negative or numerically close to zero (below $\epsilon = 0.005 = 0.5\%$) at any tenor.

\subsection{Volatility Estimation}
Practitioners discretize the tenor dimension into a finite grid $\{\tau_k\}_{k=1}^K$ and estimate the covariance matrix of forward rate changes across these tenors. Because this covariance is computed by pooling data across time, the time dimension $t$ is integrated out, yielding time-invariant volatility estimates. I use the Federal Reserve's published forward rates at annual tenor intervals for $\tau = 1,\ldots,30$ years for this volatility estimation.

I compute the sample covariance matrix of daily forward rate changes, scaled to annual units by multiplying by 252 trading days:
\begin{equation}
    \text{Cov}[\Delta f] = 252 \cdot \E[(\Delta f)(\Delta f)^\prime]
\end{equation}
where $\Delta f$ denotes the vector of daily changes in forward rates across the 30 tenor points.

Principal components analysis (PCA) is then applied to this covariance matrix to extract the dominant factors driving term structure movements. Empirically, the first three principal components typically capture a large majority of the variation in forward rate changes, corresponding to the well-known level, slope, and curvature factors. Applying eigenvalue decomposition to the covariance matrix, I extract the three eigenvectors corresponding to the three largest eigenvalues. To obtain volatility magnitudes, I scale each eigenvector by the square root of its corresponding eigenvalue, yielding adjusted volatility vectors:
\begin{equation}
    \tilde{\sigma}_n(\tau_k) = \sqrt{\lambda_n} \cdot v_n(\tau_k)
\end{equation}
where $\lambda_n$ is the $n$-th largest eigenvalue and $v_n$ is the corresponding eigenvector, evaluated at the discrete grid points $\tau_k = k$ for $k=1,\ldots,30$.

To obtain volatility functions $\sigma_n(\tau)$ defined over the continuum of tenors, I fit smooth Chebyshev polynomials through these discrete principal component loadings. Specifically, for each factor $n=1,2,3$, I fit a Chebyshev polynomial of degree 3:
\begin{equation}
    \sigma_n(\tau) = \sum_{j=0}^{3} c_{n,j} T_j\left(\frac{2\tau}{\tau_{\max}} - 1\right)
\end{equation}
where $T_j$ is the $j$-th Chebyshev polynomial of the first kind and $c_{n,j}$ are fitted coefficients.

\subsection{Computing the No-Arbitrage Drift}
A key advantage of the Svensson parameterization is that it provides an analytical expression for the derivative of the forward curve with respect to tenor:
\begin{equation}
    \frac{\d f}{\d \tau} = \left( -\frac{\beta_{1}}{\tau_{1}} + \frac{\beta_{2}}{\tau_{1}} \left( 1 - \frac{\tau}{\tau_{1}} \right) \right) \exp\!\left(-\frac{\tau}{\tau_{1}}\right)+ 	\frac{\beta_{3}}{\tau_{2}} \left( 1 - \frac{\tau}{\tau_{2}} \right) \exp\!\left(-\frac{\tau}{\tau_{2}}\right)
\end{equation}
With the volatility structure $\sigma_n(\tau)$ and forward curve derivative $\partial f/\partial \tau$ specified, the no-arbitrage drift can be computed from the Musiela condition \eqref{eq:musiela_drift}:
\begin{equation}
    \mu(t,\tau) = \frac{\partial}{\partial \tau} f(t,\tau) + \sigma(t,\tau) \cdot \int_0^\tau \sigma(t,s)\d s
\end{equation}
The second term requires evaluating the integral $\int_0^\tau \sigma(t,s)\d s$ for each volatility factor. Since the volatility functions are represented as Chebyshev polynomials, this integral is computed numerically using standard quadrature methods such as the trapezoidal rule or Simpson's rule over the discretized tenor grid $\{\tau_k\}_{k=1}^K$.

This numerical integration step is computationally inexpensive and needs to be performed only once per forward curve evaluation. Crucially, the drift computation does not require stochastic simulation—it is a deterministic calculation given the current state of the forward curve $f(t,\tau)$ and the pre-estimated volatility structure $\sigma_n(\tau)$. 

These fitted volatility functions $\sigma_n(\tau)$ and the associated drift will be held fixed during neural network training, entering the loss function through the drift term and the second-order derivative terms in the PDE.

\subsection{Local Volatility Specification}
The volatility estimation procedure described above yields time-invariant volatility functions $\sigma_n(\tau)$ that depend only on tenor. However, I adopt a local volatility specification that allows volatility to depend on the current level of forward rates. Local volatility models are widely used in interest rate derivatives markets to capture the empirical phenomenon that interest rate volatility tends to scale with the level of rates—a feature not present in constant-volatility specifications. All results presented in this paper use this local volatility structure.

The Finance-Informed Neural Network approach handles local volatility with minimal additional complexity. State-dependent volatility enters the model simply through automatic differentiation when computing the PDE residual, requiring no changes to the neural network architecture or training procedure. This flexibility stands in stark contrast to Monte Carlo methods, where local volatility substantially increases computational burden: each simulated path must now track both the forward curve evolution and the state-dependent volatility at each time step, with the volatility function evaluated thousands of times per simulation.

For the FINN implementation, the only modification required is in the data preprocessing step. I compute the covariance matrix using proportional changes rather than absolute changes:
\begin{equation}
    \text{Cov}\left[\frac{\Delta f}{\sqrt{f}}\right] = 252 \cdot \E\left[\left(\frac{\Delta f}{\sqrt{f}}\right)\left(\frac{\Delta f}{\sqrt{f}}\right)^\prime\right]
\end{equation}
where the division by $\sqrt{f}$ scales each forward rate change by the square root of the current forward rate level. After computing the scaled covariance matrix, I apply the same PCA decomposition and Chebyshev polynomial fitting described in the previous subsection. The resulting volatility functions $\sigma_n(\tau)$ now represent proportional volatilities.

During neural network training, these proportional volatilities are multiplied by $\min\{\sqrt{f(t,\tau)},M\}$ where $M = 0.4 = 40\%$ is the parameter suggested in the seminal continuous-time textbook from \textcite{shreve2004stochastic}. The cap $M$ prevents numerical instability when forward rates become extremely large—limiting the volatility scaling to reasonable levels even in high-rate environments. This choice ensures that the volatility scaling remains well-behaved across the full range of observed forward rate environments. With local volatility, the HJM dynamics under the Musiela parameterization become:
\begin{align}
     \d f(t,\tau) &= \mu(t,\tau) \d t + \sigma(t,\tau,f) \cdot \d W_t \\
     \mu(t,\tau) &= \frac{\partial}{\partial \tau} f(t,\tau) + \sigma(t,\tau,f) \cdot \int_0^\tau \sigma(t,s,f)\d s
\end{align}
where the state-dependent volatility function is defined as:
\begin{equation}
    \sigma(t,\tau,f) = \tilde{\sigma}(\tau) \cdot \min\{\sqrt{f(t,\tau)},M\}
\end{equation}
with $\tilde{\sigma}(\tau)$ denoting the PCA-derived proportional volatility functions computed during preprocessing. Importantly, the PCA-derived volatility values $\tilde \sigma_n(\tau)$ are computed only once during the preprocessing stage and stored as fixed parameters. During neural network training, multiplying these stored values by $\min\{\sqrt{f(t,\tau)},M\}$ to produce the state-dependent volatility is computationally negligible. 

With all preprocessing complete—forward curve data obtained, volatility structure estimated, and drift computation specified—the standard HJM implementation would proceed to Monte Carlo simulation for pricing path-dependent derivatives. My method avoids this computational bottleneck entirely. To explain how, I require a brief mathematical digression to the Feynman-Kac theorem, which transforms the stochastic pricing problem into a deterministic partial differential equation. Once this theoretical foundation is established, I will present the specific application to caplet pricing and describe the neural network algorithm that solves the PDE.

\section{Feynman-Kac Formula in the HJM Framework}

The Feynman-Kac theorem provides the mathematical foundation for my approach to pricing interest rate derivatives. This classical result from stochastic analysis establishes an equivalence between stochastic expectations and deterministic partial differential equations. The key insight is that instead of computing the expected discounted payoff through Monte Carlo simulation—which requires generating thousands or millions of random paths—I can solve a PDE that characterizes the same pricing functional. This section presents the multidimensional Feynman-Kac theorem and shows how to apply it to the HJM framework.


Recall the fundamental pricing formula under risk-neutral valuation. For a derivative security with contract features $\Xi$ and payoff function $h(t,f;\Xi)$ depending on the forward curve, the time-$t$ price is given by:
\begin{equation}
    V(t,\tau,f;\Xi) = \E\left[\exp\left(-\int_t^{t+\tau} r(s)\d s\right) h(t+\tau,f;\Xi) \ \Big| \ \mathcal{F}_t \right]
\end{equation}
where $\tau$ is the time-to-maturity, $r(s)=f(s,0)$ is the short rate at time $s$, $h$ characterizes the cash flow at maturity, and $\mathcal{F}_t$ is the information available at time $t$. This expectation is taken under the risk-neutral measure, reflecting the principle that the derivative price equals the present value of expected future payoffs when discounted at the risk-free rate.

Computing this expectation directly via Monte Carlo requires simulating the stochastic evolution of the entire forward curve from time $t$ to time $t+\tau$ along many paths, then averaging the discounted payoffs. The Feynman-Kac theorem transforms this stochastic problem into a deterministic PDE, avoiding the need for random simulation entirely.

\begin{theorem}[Multidimensional Discounted Feynman-Kac]
    Consider a $K$-dimensional state vector $X(t) \in \R^K$ evolving according to the stochastic differential equation:
    \begin{equation}
        \d X(t) = \mu(t,X(t))\d t + \sigma(t,X(t)) \d W_t
    \end{equation}
    where $\mu(t,X(t)) \in \R^K$ is the drift vector, $\sigma(t,X(t))\in\R^{K\times N}$ is the diffusion matrix, and $W_t\in\R^N$ is an $N$-dimensional Brownian motion. Let $r(t,X(t))$ denote a discount rate that may depend on the state.

    Define the pricing functional:
    \begin{equation}
        V(t,x) = \E^{t,x}\left[e^{-\int_t^T r(s,X(s))\d s} h(T,X(T))\right]
    \end{equation}
    where $\E^{t,x}[\cdot]$ denotes the conditional expectation given $X(t) = x$, and $h(T,X(T))$ is the terminal payoff.

    Then under general regularity conditions $V$ satisfies the partial differential equation:
    \begin{equation}
        \frac{\partial V}{\partial t} + \mu^\prime D_x V + \half \sum_{n=1}^N \sigma_n^\prime D_x^2V \sigma_n - rV = 0 \label{eq:feynman_kac_pde}
    \end{equation}
    subject to the terminal condition $V(T,x) = h(T,x)$, where $D_x V$ is the gradient vector and $D_x^2 V$ is the Hessian matrix of $V$ with respect to $x$, and $\sigma_n$ denotes the $n$-th column of the diffusion matrix.
\end{theorem}

The PDE \eqref{eq:feynman_kac_pde} has an intuitive structure. The first term $\partial V/\partial t$ captures the time evolution of the pricing functional. The second term $\mu^\prime D_x V$ represents the expected change in value due to the drift of the state variable. The third term $\frac{1}{2} \sum_{n=1}^N \sigma_n^\prime D_x^2V \sigma_n$ captures the effect of volatility through second-order (convexity) terms. Finally, the term $-rV$ discounts the value at the instantaneous rate. 


To apply the Feynman-Kac theorem to the HJM framework, I must first discretize the forward curve over the tenor dimension. The forward curve $f(t,\tau)$ is an infinite-dimensional object—a function mapping each tenor $\tau \in [0,\overline{T}]$ to a forward rate. For computational purposes, I discretize this continuum by evaluating the forward curve at a finite grid of tenor points $\{\tau_k\}_{k=1}^K$. Define the discretized forward rates as:
\begin{equation}
    f_k(t) = f(t,\tau_k) \quad \text{for } k=1,\ldots,K
\end{equation}
and collect them into a state vector $f(t) = (f_1(t), \ldots, f_K(t))^\prime \in \R^K$. This vector represents the entire forward curve at time $t$ through its values at the $K$ discrete tenor points.

Under the Musiela parameterization presented earlier, the discretized forward curve evolves according to:
\begin{equation}
    \d f(t) = \mu(t,f) \d t + \sum_{n=1}^N \sigma_n(t,f) \d W_n(t)
\end{equation}
where the drift vector and volatility factors are defined as:
\begin{align}
    \mu(t,f) &= (\mu(t,\tau_1,f), \ldots, \mu(t,\tau_K,f))^\prime \in \R^K \\
    \sigma_n(t,f) &= (\sigma_n(t,\tau_1,f), \ldots, \sigma_n(t,\tau_K,f))^\prime \in \R^K \quad \text{for } n=1,\ldots,N
\end{align}
and $\{W_n(t)\}_{n=1}^N$ are the $N$ independent Brownian motions driving term structure dynamics. This discretized system fits precisely into the framework of the multidimensional Feynman-Kac theorem with $X(t) = f(t)$ and dimension $K$.

The no-arbitrage drift $\mu(t,\tau_k,f)$ at each tenor point is computed from the Musiela condition as described in the previous section:
\begin{equation}
    \mu(t,\tau_k,f) = \frac{\partial}{\partial \tau} f(t,\tau_k) + \sigma(t,\tau_k,f) \cdot \int_0^{\tau_k} \sigma(t,s,f)\d s
\end{equation}
where for the local volatility specification, $\sigma(t,\tau_k,f) = \tilde{\sigma}(\tau_k) \cdot \min\{\sqrt{f_k(t)},M\}$. The discount rate in the Feynman-Kac theorem is the short rate $r(t) = f(t,0)$.

Applying Theorem 1 directly yields the pricing PDE for interest rate derivatives in the HJM framework:
\begin{equation}
    \frac{\partial V}{\partial t} + \mu(t,f)^\prime D_f V + \half \sum_{n=1}^N \sigma_n(t,f)^\prime D_f^2V \, \sigma_n(t,f) - r(t)V = 0 \label{eq:hjm_pde}
\end{equation}
subject to the terminal boundary condition:
\begin{equation}
    V(t+\tau,f;\Xi) = h(f;\Xi) \label{eq:hjm_boundary}
\end{equation}
where $V(t,f;\Xi)$ is the time-$t$ value of the derivative when the forward curve is $f$, and the gradient vector and Hessian matrix are defined as:
\begin{align}
    D_fV &= \left(\frac{\partial V}{\partial f_1}, \ldots, \frac{\partial V}{\partial f_K}\right)^\prime \in \R^K \\
    D_f^2V &= \left(\frac{\partial^2 V}{\partial f_k \partial f_\ell}\right)_{k,\ell=1}^K \in \R^{K \times K}
\end{align}

A crucial feature of this formulation is that the PDE \eqref{eq:hjm_pde} does \emph{not} depend on the specific contract features $\Xi$. The contract-specific information enters only through the terminal boundary condition \eqref{eq:hjm_boundary}, which specifies the payoff function $h(f;\Xi)$. This separation has important practical implications: once I have trained a neural network to solve the PDE for one contract type, adapting to a different contract requires only changing the boundary condition in the loss function. The core PDE structure—the drift term, volatility terms, and discounting—remains unchanged across all interest rate derivatives priced under the same HJM model. This modularity makes the FINN approach highly flexible for pricing a wide range of instruments. 

\section{Application: Caplet Pricing}

To demonstrate the Finance-Informed Neural Network approach, I apply it to pricing interest rate caplets. Caplets provide insurance against rising interest rates and are fundamental building blocks for more complex instruments such as interest rate caps (portfolios of caplets) and swaptions. The caplet pricing problem is particularly well-suited for demonstrating the FINN methodology because it is path-dependent in multiple ways: the payoff depends on the future LIBOR rate, which itself depends on the future forward curve, and the entire cash flow must be discounted along the stochastic path of short rates. This path dependence makes Monte Carlo simulation computationally expensive, yet the payoff structure is straightforward enough to allow clear interpretation of results.


A caplet is a call option on a future LIBOR rate. To define the contract precisely, I first introduce the LIBOR rate and then specify the caplet payoff.

The LIBOR (London Interbank Offered Rate) is an annualized interest rate for borrowing between time $\tau_1$ and time $\tau_2$.\footnote{While LIBOR has been phased out in recent years in favor of risk-free rates such as SOFR, the caplet pricing problem remains relevant for understanding interest rate derivatives more generally.} At time $t$, the forward LIBOR rate starting at $\tau_1$ and ending at $\tau_2$ is defined implicitly through the relationship between bond prices:
\begin{equation}
    L(t,\tau_1,\tau_2) = \frac{1}{\delta}\left(\frac{P(t,\tau_1)}{P(t,\tau_2)} - 1\right)
\end{equation}
where $\delta = \tau_2-\tau_1$ is the accrual period (typically 3 months or 6 months for LIBOR), and $P(t,\tau)$ is the time-$t$ price of a zero-coupon bond maturing at time $t+\tau$ with face value normalized to one. Under the HJM framework, bond prices are determined by integrating the forward curve:
\begin{equation}
    P(t,\tau) = \exp\left(-\int_0^\tau f(t,s) \d s\right)
\end{equation}
Thus the LIBOR rate depends on the entire forward curve from $0$ to $\tau_2$ through the bond price ratio, since both $P(t,\tau_1)$ and $P(t,\tau_2)$ require integrating forward rates from zero to their respective maturities.

A caplet is a call option on the LIBOR rate $L(t,\tau_1,\tau_2)$ with strike price $L_E$. The holder of the caplet receives a payoff at time $t+\tau_2$ (the end of the accrual period) equal to:
\begin{equation}
    \text{Payoff at time } t+\tau_2 = \delta \cdot \max\{L(t+\tau_1,\tau_1,\tau_2) - L_E, 0\}
\end{equation}
where the LIBOR rate is fixed at the start of the accrual period ($t+\tau_1$) and the payoff is received at the end ($t+\tau_2$). The factor $\delta$ converts the annualized rate difference into a dollar amount over the accrual period.

For computational convenience, I adopt the market convention of valuing the caplet at the settlement date $t+\tau_1$ rather than the payment date $t+\tau_2$. Discounting the payoff back one period yields the time-$(t+\tau_1)$ value:
\begin{equation}
    V(t+\tau_1,0,f) = \delta P(t+\tau_1,\delta) \max\{L(t+\tau_1,0,\delta) - L_E, 0\}
\end{equation}
where I have simplified notation by setting $\tau_1 = 0$ relative to the valuation date and $\tau_2 = \delta$. The contract features are summarized as $\Xi = (\tau,\delta,L_E)$ where $\tau$ is the time from today until the settlement date $t+\tau_1$, $\delta$ is the accrual period, and $L_E$ is the strike price.

\subsection{Neural Network Strategy}

Having derived the pricing PDE \eqref{eq:hjm_pde}, the next challenge is to solve it numerically. In principle, the PDE requires computing the pricing functional $V(t,f;\Xi)$ and its derivatives—$\partial V/\partial t$, the gradient $D_fV$, and the Hessian $D_f^2V$—at many points in the high-dimensional state space $(t,f) \in \R \times \R^K$. However, a key simplification arises from the Musiela parameterization: since the contract features $\Xi$ already include the time-to-maturity $\tau$, and the pricing problem can be solved backward from the settlement date, I can eliminate explicit dependence on calendar time $t$. The pricing functional becomes $V(\tau, f; \Xi)$ where $\tau$ represents the remaining time until settlement. Under the Musiela parameterization, since $\tau = T - t$ where $T$ is the fixed settlement date, the chain rule yields $\frac{\partial V}{\partial t} = -\frac{\partial V}{\partial \tau}$. This transforms the time derivative into a derivative with respect to time-to-maturity, allowing the neural network to be parameterized directly as a function of $(\tau, f, \Xi)$ rather than requiring separate tracking of calendar time.

Traditional finite difference methods for solving the resulting PDE become prohibitively expensive in high dimensions due to the curse of dimensionality: discretizing each dimension of the forward curve on a grid leads to exponential growth in the number of grid points. Neural networks offer a solution. A neural network is a smooth, differentiable function that can approximate complex nonlinear mappings. I parameterize the pricing function as a neural network $V_\Theta(f;\Xi)$ where $\Theta$ represents the collection of all weights and biases in the network. The network takes as inputs the discretized forward curve $f = (f_1,\ldots,f_K)$ and contract features $\Xi$ (which include time-to-maturity $\tau$), and outputs a scalar price estimate.

A key computational advantage lies in \emph{automatic differentiation}. Modern deep learning frameworks such as JAX, PyTorch, and TensorFlow implement automatic differentiation systems that can compute derivatives of any function constructed through their operations—including arbitrarily complex neural networks. Critically, these derivatives are computed exactly (up to floating-point precision), not through finite difference approximations. When I evaluate the neural network $V_\Theta(f;\Xi)$, the automatic differentiation system tracks all operations and can immediately compute the first- and second-order derivatives needed for evaluating the PDE governing the pricing functional. These derivative computations are efficient and scale easily with dimension $K$, unlike grid-based finite difference methods. An additional practical benefit is that the Greeks—sensitivities of the option price to market parameters—come essentially for free. Since the pricing PDE itself contains $\partial V/\partial \tau$ (theta, time decay) and $D_f V$ (the gradient yielding curve deltas, sensitivities to each forward rate $f_k$), these quantities are computed automatically when evaluating the PDE residual during training. Thus, evaluating pricing errors simultaneously delivers the Greeks critical for risk management and hedging strategies at zero marginal cost.

With automatic differentiation providing all necessary derivatives, I can directly evaluate how well the neural network satisfies the PDE at any point $(\tau,f)$. The PDE residual at a point is:
\begin{equation}
     -\frac{\partial V_\Theta}{\partial \tau} + \mu(t,f)^\prime D_f V_\Theta + \half \sum_{n=1}^N \sigma_n(t,f)^\prime D_f^2V_\Theta \, \sigma_n(t,f) - r(t)V_\Theta
\end{equation}

\subsection{FINN Architecture and Training}

For the remainder of this section, I adopt the notation $\mathcal X$ to denote the inputs to the neural network and $\mathcal Y$ to denote the outputs. The inputs are $\mathcal X = (f_1, \ldots, f_K, S, \Xi)\in\R^{K+9}$ comprising the discretized forward curve, Svensson parameters $S = (\beta_0, \beta_1, \beta_2, \beta_3, \tau_1, \tau_2)\in\R^{6}$, and contract features $\Xi = (\tau, \delta, L_E)\in\R^{3}$ where $\tau$ is the time until the start of the accrual period, $\delta$ is the accrual length, and $L_E$ is the strike price. Note that $\tau_2 = \tau + \delta$ is redundant and not included as a separate input. The output is the scalar caplet price $\mathcal Y = V(\tau,f;\Xi)$.

To improve training stability and convergence, I normalize inputs within the network architecture itself—crucially, as the first layer before any nonlinear transformations. This normalization is implemented as a differentiable operation within the computational graph, meaning automatic differentiation seamlessly handles all derivative computations through the normalization without requiring manual adjustment. The time-related contract features $\tau_1$ and $\delta$ are normalized by dividing by the maximum maturity $\tau_{\max}$ in the data. The Svensson parameters are transformed to z-scores:
\begin{equation}
    \tilde{\beta}_i = \frac{\beta_i - \mu_{\beta_i}}{\sigma_{\beta_i}}
\end{equation}
where $\mu_{\beta_i}$ and $\sigma_{\beta_i}$ are the empirical mean and standard deviation of parameter $\beta_i$ computed across the historical dataset. The forward rates themselves are left unnormalized, as their scale is economically meaningful and directly enters the pricing formula.

The FINN architecture consists of three hidden layers, each with 500 neurons. 
I apply the sigmoid linear unit (SiLU) activation function $\text{silu}(x) = x\times\sigma(x)$ where $\sigma(x)=\frac{1}{1+\exp(-x)}$ is the sigmoid function to the hidden layers, which provides smooth, differentiable nonlinearity necessary for backpropagation through the financial equilibrium conditions. 
Since prices must be nonnegative, I apply the softplus activation function $\text{softplus}(x) = \log(1 + e^x)$ to the output layer, which smoothly enforces the nonnegativity constraint while remaining differentiable everywhere. 

To accelerate training, I precompute several quantities that remain constant across training iterations. First, I enumerate all admissible tenor-accrual pairs $(\tau, \delta)$, storing these as indices for rapid sampling during batch generation. Second, since all forward rates and volatility evaluations occur on the fixed tenor grid $\{\tau_k\}_{k=1}^K$, I precompute the trapezoidal integration matrix $C \in \R^{K \times K}$ where element $C_{ij}$ gives the weight for integrating from grid point $j$ to grid point $i$. This vectorizes all integral computations—including bond price calculations $P(t,\tau) = \exp(-\int_0^\tau f(t,s) \d s)$ and drift integrals—into matrix-vector products. Third, I precompute and store the Chebyshev polynomial coefficients for the volatility functions $\sigma_n(\tau)$ from the PCA decomposition. For the constant volatility specification, I additionally precompute the volatility integral $\int_0^\tau \sigma(t,s) \d s$ needed in the drift term, storing these as vectors that can be directly used in the loss function evaluation. These precomputations transform computationally expensive operations into simple lookups and matrix multiplications, substantially reducing per-iteration training time.

Unlike traditional supervised learning where training data comes from a pre-existing dataset, the FINN approach generates its own training data by sampling from the historical forward curve distribution. Each training batch is constructed by randomly sampling:
\begin{itemize}
    \item[-] A tenor-accrual pair with $\tau \in [0,5]$ and $\delta \in [1/3,3/4]$ (covering typical caplet accrual periods and maturities)
    \item[-] A forward curve $f$ and its corresponding Svensson parameters from the filtered historical dataset
    \item[-] A strike price $L_E$ from a Chebyshev grid over $[0, 0.07]$ (covering the range of observed forward rates)
\end{itemize}
This sampling strategy ensures the network trains on a diverse set of market conditions, contract specifications, and moneyness levels. The training data is regenerated every epoch, preventing overfitting to any particular set of forward curves and encouraging the network to learn the underlying PDE structure rather than memorizing specific curve realizations.  

To evaluate the boundary condition, for any historically observed forward curve $f$ sampled from the data, I simply set the time-to-maturity $\tau = 0$ in the contract features and evaluate what the payoff would be if that curve represented the state at expiry. That is, I treat each sampled historical curve as if it were the forward curve at the settlement date and compute the corresponding caplet payoff. The loss function then penalizes deviations between the network's prediction $V_\Theta(\tau=0,f;\Xi)$ and this analytically computed payoff. This approach leverages the rich historical variation in forward curve shapes to train the boundary condition, while the PDE loss teaches the network how values at earlier times $\tau > 0$ relate to these terminal payoffs through the no-arbitrage dynamics. Importantly, this strategy completely avoids the computational expense of path simulation—all training occurs on the cross-section of historical curves, not their time-series evolution.

To further enrich the disciplining of the pricing functional, I exploit the special case of zero-strike caplets, which admit a closed-form analytical expression that holds at all times, not just at maturity. For a caplet with strike $L_E = 0$, there is no optionality—the payoff is deterministically equal to the LIBOR rate. Standard no-arbitrage arguments yield the closed-form expression:
\begin{equation}
    V(\tau_1,f;L_E=0) = P(\tau_1) - P(\tau_1+\delta) = \exp\left(-\int_0^{\tau_1} f(s)\d s\right) - \exp\left(-\int_0^{\tau_1+\delta} f(s)\d s\right)
\end{equation}
Crucially, this closed-form expression holds for any $\tau_1 \geq 0$, not just at the boundary. The absence of optionality means the pricing PDE reduces to a deterministic bond pricing formula, which can be evaluated exactly from the current forward curve. During training, I augment the loss function by relabeling a fraction of the sampled data points to have strike $L_E=0$ at various times-to-maturity $\tau_1$ and penalizing deviations between the network's zero-strike prediction and this analytical formula. This provides the network with exact supervisory signals across the full domain—both in the time dimension and across the space of forward curve shapes—serving as a powerful regularization that anchors the pricing functional to known analytical values. The zero-strike case effectively provides thousands of additional training points where the true solution is known exactly, complementing the PDE residual minimization and boundary condition training for positive strikes, thereby improving overall accuracy and accelerating convergence.

The network is trained using the Adam optimizer \parencite{kingma2014adam} with weight decay regularization ($\lambda = 10^{-5}$) in a three-regime curriculum learning schedule. Each regime uses progressively smaller learning rates to refine the solution:
\begin{itemize}
    \item[-] {Regime 1}: 15,000 epochs with learning rate $10^{-4}$, batch size 100, 10 batches per epoch
    \item[-] {Regime 2}: 5,000 epochs with learning rate $10^{-5}$, batch size 100, 10 batches per epoch
    \item[-] {Regime 3}: 2,500 epochs with learning rate $10^{-6}$, batch size 500, 2 batches per epoch
\end{itemize}
This curriculum learning approach starts with larger learning rates to explore the solution space broadly, then progressively refines with smaller learning rates and larger batch sizes. Crucially, the training data is redrawn after each epoch, ensuring that the network learns from fresh forward curve realizations rather than overfitting to a single sample.

A key computational challenge in evaluating the PDE residual is computing the second-order term
, which involves the Hessian matrix $D_f^2V \in \R^{K \times K}$. 
I employ a directional derivative trick to avoid materializing the full Hessian. Using the chain rule identity, the quadratic form can be rewritten as:
\begin{equation}
    \sigma_n^\prime D_f^2V \sigma_n = D_f(D_f V \cdot \sigma_n) \cdot \sigma_n
\end{equation}
The algorithm proceeds as follows: first compute the gradient $D_f V$ once. Then for each factor $n$, form the scalar directional derivative $s = D_f V \cdot \sigma_n$, compute its gradient $D_f s$ (which gives the second derivative in the $\sigma_n$ direction), and dot with $\sigma_n$ again to obtain $\sigma_n^\prime D_f^2V \sigma_n$. 
This approach reduces memory and computational requirements, making it feasible to evaluate the PDE residual efficiently even for large discretizations $K$.

Combining these elements, the complete training objective becomes clear. If $V_\Theta$ were the true solution, this residual would equal zero everywhere. I train the neural network to minimize the squared PDE residual across a collection of randomly sampled points, combined with penalties for violating both the terminal boundary condition and the zero-strike analytical formula. The Finance-Informed loss function comprises three components, averaged over number of contracts $N_\text{batch}$:
\begin{multline}
    \mathcal L_\Theta = \frac{1}{N_\text{batch}}\sum_{n=1}^{N_\text{batch}} \Bigg(\left(-\frac{\partial V_\Theta}{\partial \tau_1} + \mu^\prime D_f V_\Theta + \half \sum_{n=1}^N \sigma_n^\prime D_f^2V_\Theta \sigma_n - r V_\Theta\right)^2 \\
    + \left(V_\Theta(\tau_1=0,f;\Xi) -  \delta P(\delta) \max\{L(0,\delta) - L_E, 0\}\right)^2 \\
    + \left(V_\Theta(\tau_1,f;L_E=0) - (P(\tau_1) - P(\tau_1+\delta))\right)^2\Bigg)
\end{multline}
where the first term penalizes violations of the pricing PDE across sampled interior points $(\tau_1, f)$ with $\tau_1 > 0$, the second term enforces the terminal boundary condition at settlement ($\tau_1=0$) for positive-strike caplets, and the third term anchors the network to the analytical zero-strike formula across all times-to-maturity. The neural network parameters are updated through standard gradient descent (specifically, the Adam optimizer) by solving:
\begin{equation}
    \Theta^* = \argmin{\Theta}{\mathcal L_\Theta}
\end{equation}

This approach—using automatic differentiation to enforce PDE structure through a loss function—allows the neural network to learn the pricing function without ever simulating Monte Carlo paths. The method scales efficiently to high-dimensional state spaces and handles path-dependent payoffs through the PDE framework rather than through stochastic simulation.

Table \ref{tab:computationalCH3} reports the hardware used in training the model. Since the FINN generates its own training data rather than requiring large pre-existing datasets, the memory bottleneck is the network size rather than dataset size. This makes consumer-grade hardware with modest GPU memory (8GB VRAM) entirely sufficient, eliminating the need for specialized high-performance computing infrastructure typical of large-scale machine learning applications.
\begin{table}[h]
\centering
\caption{Computational Environment}
\label{tab:computationalCH3}
\begin{tabular}{ll}
\toprule
\textbf{Hardware} & \\
\midrule
Processor & 12th Gen Intel i9-12900KF (24) @ 5.100GHz \\
GPU & NVIDIA GeForce RTX 3080 with 8GB of VRAM \\
RAM & 32 GB \\
\bottomrule
\end{tabular}
\end{table}

To assess how solution accuracy and computational cost scale with the dimension of the discretized forward curve, I train separate models for eight different discretizations, varying the parameter $K \in \{10, 25, 35, 50, 75, 100, 125, 150\}$. Remarkably, training time remains approximately one hour across all discretization levels, demonstrating that the FINN approach scales gracefully with dimension—a stark contrast to traditional finite difference or Monte Carlo methods where computational cost grows exponentially or linearly (respectively) with $K$. This dimensional robustness arises from the efficiency of automatic differentiation and the fact that neural network evaluation cost grows only modestly with input dimension.

\subsection{Results}

I validate the FINN approach by comparing caplet prices to Monte Carlo benchmarks across all eight discretization levels $K \in \{10, 25, 35, 50, 75, 100, 125, 150\}$. All results employ the local volatility specification described above, where volatility scales with the square root of the forward rate level. For each model, I generate a test set of 1,000 randomly sampled forward curves with varying caplet contract specifications (strikes, tenors, and accrual periods) and compute both FINN prices and Monte Carlo prices using 10,000 simulated paths per contract. The Monte Carlo benchmark likewise uses the same local volatility structure, ensuring a consistent comparison. The results demonstrate that FINNs achieve comparable pricing accuracy to Monte Carlo simulation while delivering transformative computational speedups.

\begin{figure}[!htb]
\centering
\includegraphics[width=\linewidth]{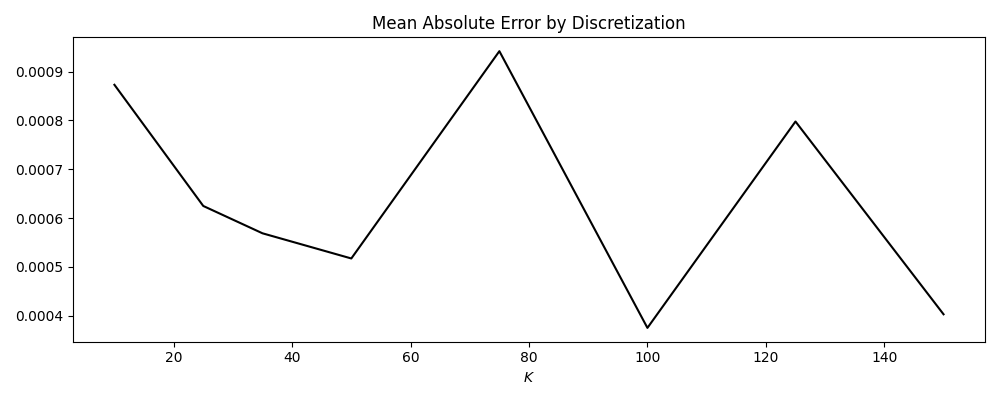}
\caption{Mean Absolute Pricing Error vs. Discretization Level (1000-contract test set). All discretizations maintain error below 0.001, well within acceptable bounds for trading applications. The finest discretizations ($K=100$ and $K=150$) achieve the tightest error bounds around 0.0004. The non-monotonicity suggests complex interactions between discretization level, training dynamics, and network capacity.}
\label{fig:error}
\end{figure}

Figure \ref{fig:error} quantifies pricing accuracy by plotting the mean absolute error between FINN and Monte Carlo prices as a function of $K$. The error profile does not decrease monotonically with $K$—instead, it exhibits a non-monotonic pattern with local minima around $K=50$ and $K=100$, and local peaks around $K=10$ and $K=75$. All errors remain below 0.001 (0.1 cents on a dollar-denominated contract), with most discretizations achieving errors between 0.0004 and 0.0007. The finest discretizations ($K=100$ and $K=150$) both achieve errors around 0.0004, representing approximately 0.04 cents per dollar of contract value. This non-monotonicity suggests that discretization level interacts with the training dynamics and neural network capacity in complex ways, rather than finer grids uniformly improving accuracy. For practical purposes, all discretization levels achieve sufficient accuracy for trading applications, with $K=100$ and $K=150$ providing the tightest error bounds.

An important caveat: Monte Carlo prices themselves are not exact benchmarks but rather approximations subject to their own sources of error. The Monte Carlo implementation discretizes the forward curve into $K$ tenor points (the same discretization used by the FINN), discretizes time into finite steps (Euler-Maruyama Markov chain), and performs repeated numerical integrations to produce drift terms. Crucially, when computing the drift term, Monte Carlo must approximate the tenor derivative $\partial f/\partial \tau$ using finite differences on the discretized grid, introducing additional discretization error. By contrast, the FINN leverages the analytical Svensson formula for $\partial f/\partial \tau$, avoiding this source of approximation entirely. Additionally, Monte Carlo prices are subject to sampling error despite using 10,000 paths per contract. The reported errors therefore reflect the combined approximation errors from both methods rather than pure FINN error relative to a known analytical solution. In principle, the FINN could be closer to the true price than the Monte Carlo benchmark, particularly if the neural network captures the smooth PDE solution more accurately than discrete-time simulation with finite paths. The sub-0.001 agreement between methods suggests both approaches achieve high accuracy, but neither should be viewed as providing ground truth.

\begin{figure}[!htb]
\centering
\includegraphics[width=\linewidth]{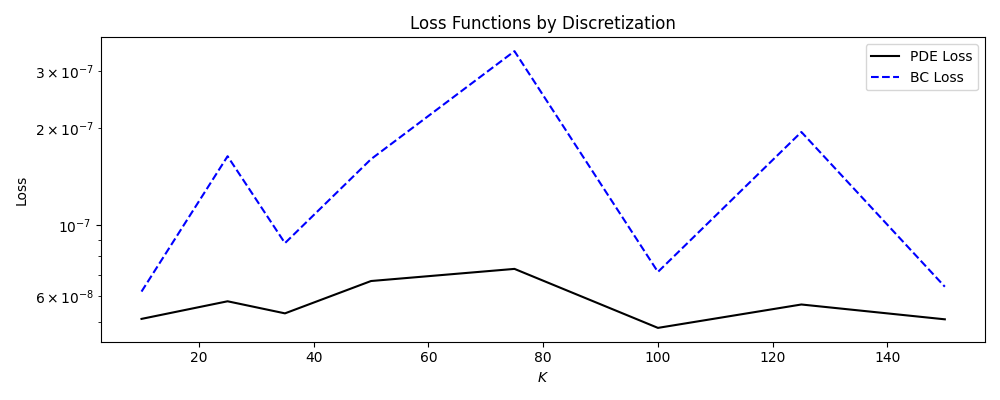}
\caption{Final Training Loss Across Discretizations. Both PDE residual (solid black) and boundary condition violation (dashed blue) remain stable across all values of $K$, with losses on the order of $10^{-7}$ to $10^{-8}$. The consistency of loss magnitudes demonstrates that the FINN training procedure scales robustly with dimension.}
\label{fig:losses}
\end{figure}

Figure \ref{fig:losses} displays the final PDE loss (solid black line) and boundary condition (BC) loss (dashed blue line) achieved after training for each discretization level. Both loss components remain remarkably stable across $K$, with PDE loss fluctuating around $5 \times 10^{-8}$ and BC loss varying between approximately $5 \times 10^{-8}$ and $4 \times 10^{-7}$. The y-axis uses a logarithmic scale, highlighting that final losses are uniformly small (on the order of $10^{-7}$ to $10^{-8}$), indicating high-precision satisfaction of both the pricing PDE and terminal boundary conditions across all discretization levels. The consistency of these loss values confirms that the FINN training procedure scales robustly with dimension—there is no evidence of training degradation at higher $K$.

\begin{figure}
\centering
\includegraphics[width=.8\linewidth]{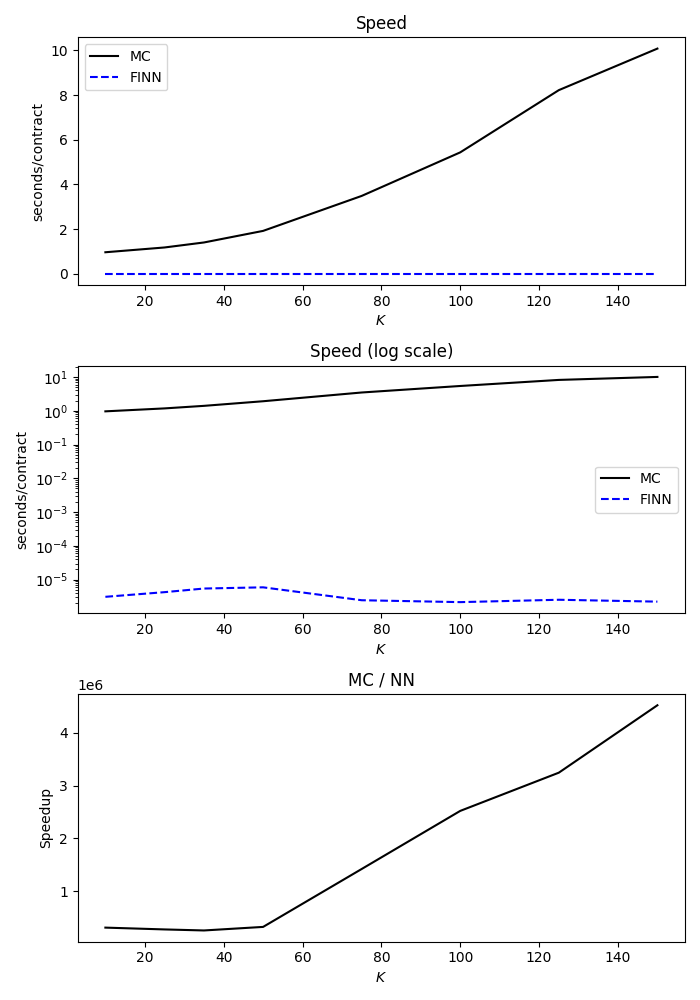}
\caption{Pricing Speed: FINNs vs. Monte Carlo (1000-contract test set). Top panel: Monte Carlo time (solid) grows linearly from 1 to 10 seconds per contract as $K$ increases, while FINN time (dashed) remains near zero. Middle panel (log scale): FINN evaluation takes $\sim10^{-6}$ to $10^{-5}$ seconds (a few microseconds), while MC takes 1 to 10 seconds. Bottom panel: Speedup ratio grows from approximately 300,000× at $K=10$ to over 4.5 million× at $K=150$, representing a multi-million-fold computational advantage that increases with dimension.}
\label{fig:speed}
\end{figure}

The most compelling advantage of the FINN approach lies in computational speed. Figure \ref{fig:speed} presents comprehensive comparisons of FINN and Monte Carlo evaluation times. The top panel of Figure \ref{fig:speed} shows pricing time in seconds per contract on a linear scale: Monte Carlo time (solid black line) grows from approximately 1 second at $K=10$ to nearly 10 seconds at $K=150$, while FINN time (dashed blue line) remains essentially flat near zero across all discretizations. The middle panel displays the same data on a log scale, revealing that FINN evaluation takes approximately $10^{-6}$ to $10^{-5}$ seconds per contract (a few microseconds), while Monte Carlo ranges from $1$ to $10$ seconds. The bottom panel shows the speedup ratio (MC time / FINN time), which grows from approximately 300,000 at $K=10$ to over 4.5 million at $K=150$—representing a \emph{hundred-thousand to multi-million-fold speedup}. Table \ref{tab:timing} provides the precise timing measurements and mean absolute errors for each discretization level.

This is the headline result: FINNs price caplets roughly \emph{300,000 to 4,500,000 times faster} than Monte Carlo simulation, with the advantage increasing as dimension grows. Once trained, the neural network evaluates in a few microseconds regardless of the forward curve dimensionality, while Monte Carlo evaluation time grows linearly with $K$. This dramatic speed advantage makes FINNs transformative for applications requiring rapid repricing of large portfolios, such as real-time risk management, high-frequency trading, or iterative calibration procedures.

\begin{table}[h]
\centering
\begin{tabular}{c|ccc|c}
\hline
$K$ & MC Time & FINN Time & Speedup Multiple & MAE \\
\hline
$10$ & $0.96$ & $3.10 \times 10^{-6}$ & $311{,}366$ & $8.73 \times 10^{-4}$ \\
$25$ & $1.18$ & $4.25 \times 10^{-6}$ & $276{,}943$ & $6.25 \times 10^{-4}$ \\
$35$ & $1.40$ & $5.45 \times 10^{-6}$ & $257{,}440$ & $5.69 \times 10^{-4}$ \\
$50$ & $1.92$ & $5.91 \times 10^{-6}$ & $324{,}690$ & $5.17 \times 10^{-4}$ \\
$75$ & $3.49$ & $2.45 \times 10^{-6}$ & $1{,}421{,}537$ & $9.42 \times 10^{-4}$ \\
$100$ & $5.44$ & $2.16 \times 10^{-6}$ & $2{,}522{,}035$ & $3.75 \times 10^{-4}$ \\
$125$ & $8.22$ & $2.53 \times 10^{-6}$ & $3{,}245{,}742$ & $7.98 \times 10^{-4}$ \\
$150$ & $10.08$ & $2.23 \times 10^{-6}$ & $4{,}522{,}638$ & $4.03 \times 10^{-4}$ \\
\hline
\end{tabular}
\caption{{Computational performance and accuracy across discretization phases. MC Time and FINN Time are in seconds per contract. Speedup Multiple is the ratio of MC Time to FINN Time. MAE (Mean Absolute Error) measures the difference between FINN and Monte Carlo prices.}}
\label{{tab:timing}}
\end{table}

\begin{figure}[H]
\centering
\includegraphics[width=.9\linewidth]{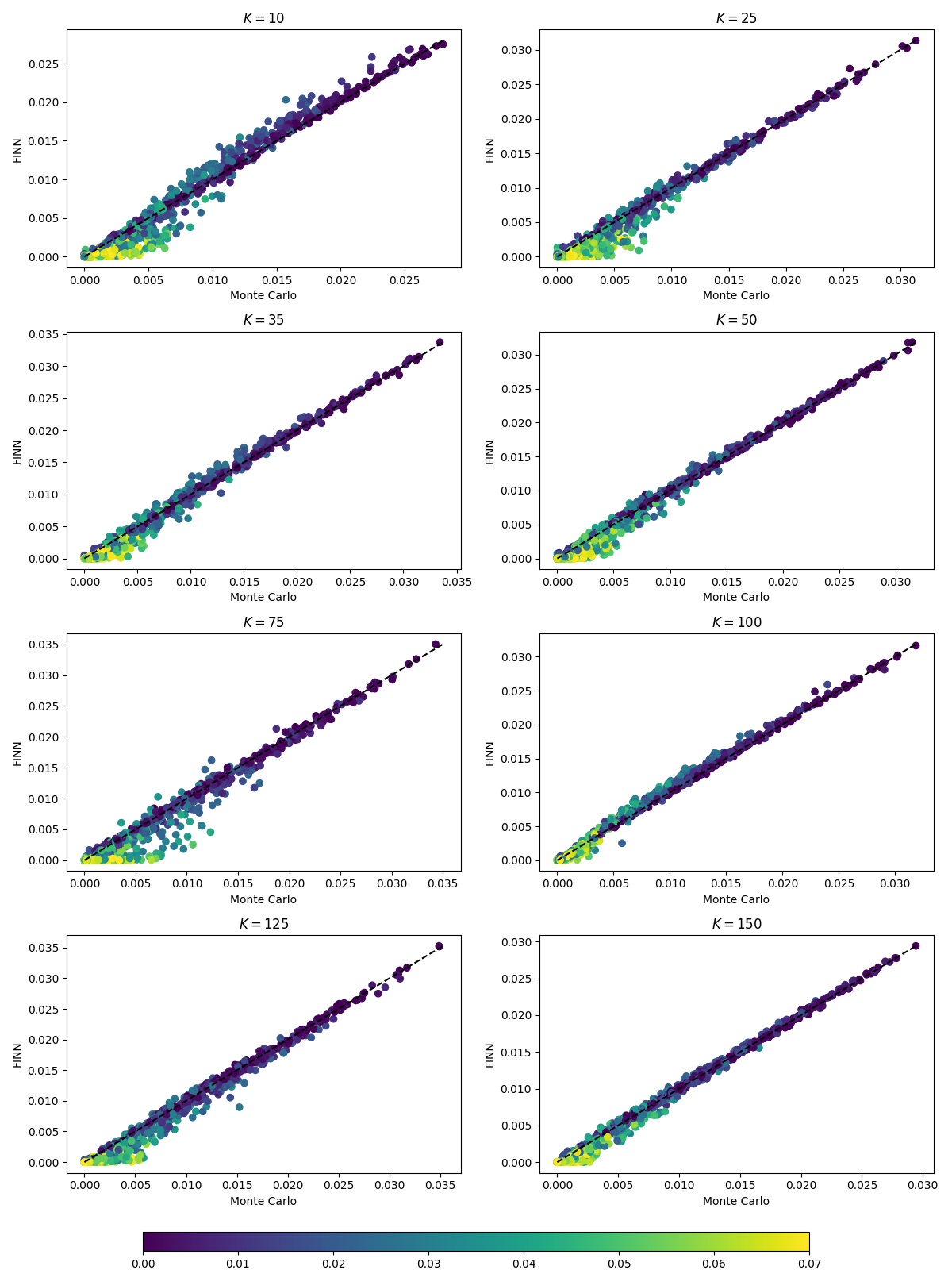}
\caption{FINN vs. Monte Carlo Prices Across Discretizations (1000-contract test set). Each panel shows test contracts for a different discretization level $K$, with points colored by strike price $L_E$. Points cluster tightly along the 45-degree line, with near-zero strike contracts (purple) showing the tightest agreement due to the analytical zero-strike anchoring. Higher-strike contracts (yellow/green) exhibit slightly more scatter, suggesting the FINN performs best near the analytical anchor and pointing to future work on dual anchoring from both extremes.}
\label{fig:mcnn_scatter}
\end{figure}

Figure \ref{fig:mcnn_scatter} plots FINN prices against Monte Carlo prices for all eight discretization levels. Each panel corresponds to a different value of $K$, with points colored by strike price—purple indicates near-zero strikes (high caplet prices near the top-right of each panel), while yellow indicates high strikes near 7\% (lower prices, deep out-of-the-money contracts near the origin). The scatter plots reveal tight clustering along the 45-degree line across all discretizations, demonstrating strong agreement between FINN and Monte Carlo prices. Notably, the near-zero strike contracts (purple points with highest prices) exhibit particularly tight agreement, reflecting the zero-strike analytical anchoring used during training. However, higher-strike contracts (yellow/green points at lower prices, farther from the zero-strike anchor) show slightly more scatter, particularly visible in panels with coarser discretizations. This suggests that the FINN struggles more with contracts farther from the analytical anchor point, pointing to potential future work: anchoring from both the zero-strike and high-strike extremes could improve accuracy across the full moneyness spectrum.

\section{Conclusion}

This paper demonstrates a fundamental computational breakthrough for pricing path-dependent interest rate derivatives under the Heath-Jarrow-Morton framework. The HJM model offers unparalleled generality: it specifies arbitrage-free dynamics for the entire forward curve simultaneously, nesting earlier short-rate models as special cases and providing a theoretically rigorous foundation for derivatives pricing. However, this generality has historically come at a severe computational cost. The forward curve is infinite-dimensional, and pricing path-dependent contracts traditionally requires Monte Carlo simulation of thousands or millions of stochastic paths—a computational burden that grows linearly with the dimensionality of the discretized state space.

I circumvent Monte Carlo simulation entirely by invoking the Feynman-Kac theorem, which establishes that stochastic expectations can be characterized as solutions to deterministic partial differential equations. Rather than simulating random forward curve paths to estimate expected payoffs, I solve the PDE governing the pricing functional directly. This transformation eliminates the need for path generation, replacing stochastic simulation with a deterministic PDE plus boundary value problem. However, this substitution merely shifts the computational challenge: traditional finite difference and finite element methods for solving high-dimensional PDEs suffer from the curse of dimensionality, with memory and computational requirements growing exponentially in the number of state variables.

The solution lies in Finance-Informed Neural Networks (FINNs)—deep learning models trained to satisfy the pricing PDE by embedding the differential equation directly into the loss function. The neural network parameterizes the pricing functional, and automatic differentiation computes the exact derivatives needed to evaluate PDE residuals at any point in the state space. Training minimizes violations of the governing PDE across sampled forward curve realizations, combined with penalties for violating terminal boundary conditions and deviations from analytical zero-strike solutions. This approach leverages three key advantages of neural networks: universal approximation capability for complex nonlinear functions, smooth differentiability enabling automatic computation of all required derivatives, and graceful scaling with input dimension.

Crucially, FINN evaluation cost does not grow with the size of the state space. Once trained, the neural network prices derivatives nearly instantaneously regardless of whether the discretized forward curve contains ten tenor points or one hundred and fifty. Monte Carlo methods, by contrast, exhibit computational cost that grows linearly with discretization level—each additional state variable requires simulating more data and more computational expense in the form of integrals. This dimensional robustness makes FINNs transformative for pricing derivatives in high-dimensional settings where traditional methods become prohibitively expensive.

An additional practical advantage emerges from the structure of the pricing PDE itself: the major Greeks—theta (time decay) and curve deltas (sensitivities to each forward rate)—appear directly as terms in the differential equation. Since automatic differentiation computes these quantities when evaluating the PDE residual during training, they are obtained at zero marginal cost once the network is trained. Other Greeks, such as gamma (convexity) and vega (volatility sensitivity), require computing additional derivatives but remain computationally cheap via the same automatic differentiation framework. This stands in stark contrast to Monte Carlo methods, which cannot provide Greeks without complete re-simulation. To compute a single delta via Monte Carlo requires perturbing the contract parameter and running thousands of paths anew—and because even small parameter perturbations can significantly alter simulated prices through the stochastic dynamics, each Greek calculation demands a fresh Monte Carlo simulation. For a portfolio requiring hundreds of prices and thousands of sensitivities, this quickly becomes computationally prohibitive. Once the FINN is trained, however, all prices and all Greeks are available functionally and instantaneously through simple forward and backward passes of the neural network, with automatic differentiation delivering exact derivatives at negligible cost. For practitioners managing large derivatives portfolios, this represents a fundamental improvement over Monte Carlo: the ability to compute thousands of prices and their complete associated risk profiles nearly instantaneously, without any re-simulation.

The empirical results validate this approach decisively. Tested on 1,000 randomly sampled caplet contracts across eight discretization levels ($K \in \{10, 25, 35, 50, 75, 100, 125, 150\}$), FINNs achieve pricing accuracy within 0.04 to 0.07 cents per dollar of contract value compared to Monte Carlo benchmarks. These errors are well within acceptable bounds for trading applications, particularly considering that the Monte Carlo benchmark itself is subject to multiple sources of approximation error: discretization of the forward curve into $K$ nodes, discrete time-stepping via Euler-Maruyama, sampling error despite using 10,000 paths, and crucially, approximation of the tenor derivative $\partial f/\partial \tau$ via finite differences on the discretized grid. The FINN, by contrast, uses the analytical Svensson formula for this derivative, avoiding this source of discretization error. Given that both methods involve approximations, the sub-0.1 cent agreement between FINN and Monte Carlo prices demonstrates that FINNs achieve pricing accuracy comparable to—and possibly exceeding—traditional simulation methods. The computational advantage is dramatic: FINNs price caplets 300,000 to 4,500,000 times faster than Monte Carlo simulation, with speedups increasing as dimension grows. Once trained on consumer-grade hardware (8GB GPU), evaluation takes only a few microseconds per contract regardless of discretization level. Monte Carlo evaluation time, by contrast, grows linearly with $K$, reaching nearly 10 seconds per contract at $K=150$. This scaling behavior makes FINNs particularly attractive for real-time risk management, high-frequency trading, iterative model calibration, and any application requiring rapid repricing of large derivatives portfolios under varying market conditions.

Beyond the immediate application to caplet pricing, the FINN methodology generalizes naturally to other path-dependent interest rate derivatives—caps, floors, swaptions, callable bonds—all of which can be priced by modifying only the boundary condition in the loss function while retaining the same core PDE structure. The framework already accommodates local volatility (all results in this paper use volatility that scales with the square root of forward rate levels), and extends equally to more sophisticated stochastic volatility models and jump-diffusion processes, with state-dependent coefficients entering seamlessly through automatic differentiation. This flexibility, combined with the dramatic computational speedups and essentially free Greeks, positions Finance-Informed Neural Networks as a powerful new tool for derivatives pricing in high-dimensional continuous-time models.

\newpage
\addcontentsline{toc}{section}{References}
{\emergencystretch=1em \printbibliography}

\end{document}